\documentclass[10pt,conference,onecolumn]{IEEEtran}
\usepackage{amssymb,amsmath,epsfig,cite,graphicx,ltablex,tabularx,setspace,caption}
\usepackage{multirow}

\begin{document}

\title{Speech Polarity Detection Using Hilbert Phase Information}

\author{\IEEEauthorblockN{D Govind, Anju Susan Biju and Aguthu Smily}
\IEEEauthorblockA{Center for Computational Engineering and Networking\\
Amrita Vishwa Vidyapeetham (University), Coimbatore, Tamilnadu\\
Email: d\_govind@cb.amrita.edu,\{smily.agthu,anjusb118\}@gmail.com}}

\maketitle

\begin{abstract}
The objective of the present work is to propose a method to automatically detect polarity of the speech signals by estimating instants of significant excitation of the vocaltract and the cosine phase of the analytic signal representation. The phase changes in the analytic signal around the Hilbert envelope (HE) peaks are found to vary according to the polarity of the given speech signal. The relevant HE peaks for the Hilbert phase analysis are selected by estimating the instants of significant excitation in speech. The speech polarity identification rate obtained for the proposed method is almost equal to the state of the art residual skewness method for speech polarity detection. The proposed method  also provides the same results for the polarity detection in electro-glottogram signals. Finally, the robustness of the proposed method is confirmed from the reduced detection error rates obtained in noisy environments with various signal to noise ratios (SNRs). The MATLAB codes used for implementing the proposed method are available for download from the following link: $http://nlp.amrita.edu:8080/TTS/polarityprograms.zip$
\end{abstract}
\section{Introduction}

The change in the polarity of the electrical connections of the speech recording devices is the root cause for the polarity inversion in the recorded speech ~\cite{Dugman2011}~\cite{Ding1998}.  The polarity inversion in speech signals are perceptually less significant. However, the polarity inversion reduces the performance of the speech parameter estimation algorithms ~\cite{Dugman2011}. For instance, in case of pitch estimation, a significant difference in the pitch values can be observed for the same speech signals with opposite polarities. These variations in the estimated pitch values may create discontinuities at the concatenation points of the units selected from the speech signals recorded with opposite polarities which in turn reduce the perceptual quality of the synthesized speech ~\cite{Ding 1998}.  Hence polarity detection is an essential component which has to be performed after the speech recording to correct the polarity of the speech to guarantee the performance of the popular speech parameter estimation algorithms.

There are various works reported in the literature to automatically detect the polarity of the speech signals~\cite{Dugman2011, Ding1998, Drugman2013, Saratxaga2009}. Ding et al., proposed a method based on the spurious glottal wave derived from the linear prediction (LP) residual of speech for polarity detection~\cite{Ding1998}.  Drugman et al. used phase shifts of the oscillating moments around the local fundamental frequencies as an indication of the polarity of the speech~\cite{Dugman2011}. Based on the Phase analysis of the harmonics in speech, Saratxaga et al. introduced phase cut and relative phase shifts (RPS) for polarity detection~\cite{Saratxaga2009}. Drugman, in his recent work, proposed a robust speech polarity detection algorithm by computing the statistical skewness between LP residual of speech and its asymmetric version of LP residual (rough approximation of the glottal source)~\cite{Drugman2013}. Sign of the skewness value indicates the polarity of the speech. This residual skewness based polarity detection (RESKEW) is the best existing approach for the automatic determination of the speech polarity. Even though , RESKEW is proposed for  polarity detection in speech signals, the method is found to be working well for electro-glottogram (EGG) signals also.

The present work aims at extracting the polarity information by analyzing the phase variations occurring around the events correspond to instants of significant excitation in speech signals.  Instants of significant excitation are referred to the events in speech at which the excitation of the vocaltract is maximum~\cite{Murty2008}. The polarity is determined from the sign of the slopes in the cosine phase around the zero crossings correspond to Hilbert envelope (HE) peaks nearest to the estimated instants. The organization of the paper is as follows: The computation of cosine phase from the analytic signal is given in Section~\ref{compute_analytic}. The proposed algorithm for polarity detection using Hilbert transform and glottal closure instants are provided in Section~\ref{proposed}. Section~\ref{exp_results} describes the experimental performance evaluation. Section~\ref{summary} summarizes the work.

\section{Computation of Cosine Phase Using Hilbert transform}
\label{compute_analytic}
The cosine phase of the analytic signal representation of the given original speech $s(n)$ is computed as in Equation~\ref{cosine_phase},

\begin{equation}
\label{cosine_phase}
cos(\phi(n))=\frac{s(n)}{\sqrt{s^2(n)+s_h^{2}(n)}}
\end{equation}

Where, $s_h(n)$ is the Hilbert transform of $s(n)$ and the term $\sqrt{s^2(n)+s_h^{2}(n)}$ is the expression for HE
 which is defined as the magnitude of the complex analytic signal,$s_a(n)$, expressed as given by the Equation~\ref{analytic_signal},
\begin{equation}
\label{analytic_signal}
s_a(n)=s(n)+j.s_h(n)
\end{equation}

$s_a(n)$ is the complex time signal having original speech signal, $s(n)$, as the real part and the Hilbert transformed speech $s_h(n)$ as the imaginary part~\cite{Cohen1995}. \\

\section{Methodology}
\label{proposed}

\begin{figure*}
\label{HPFSpeechHE}
\epsfig{figure=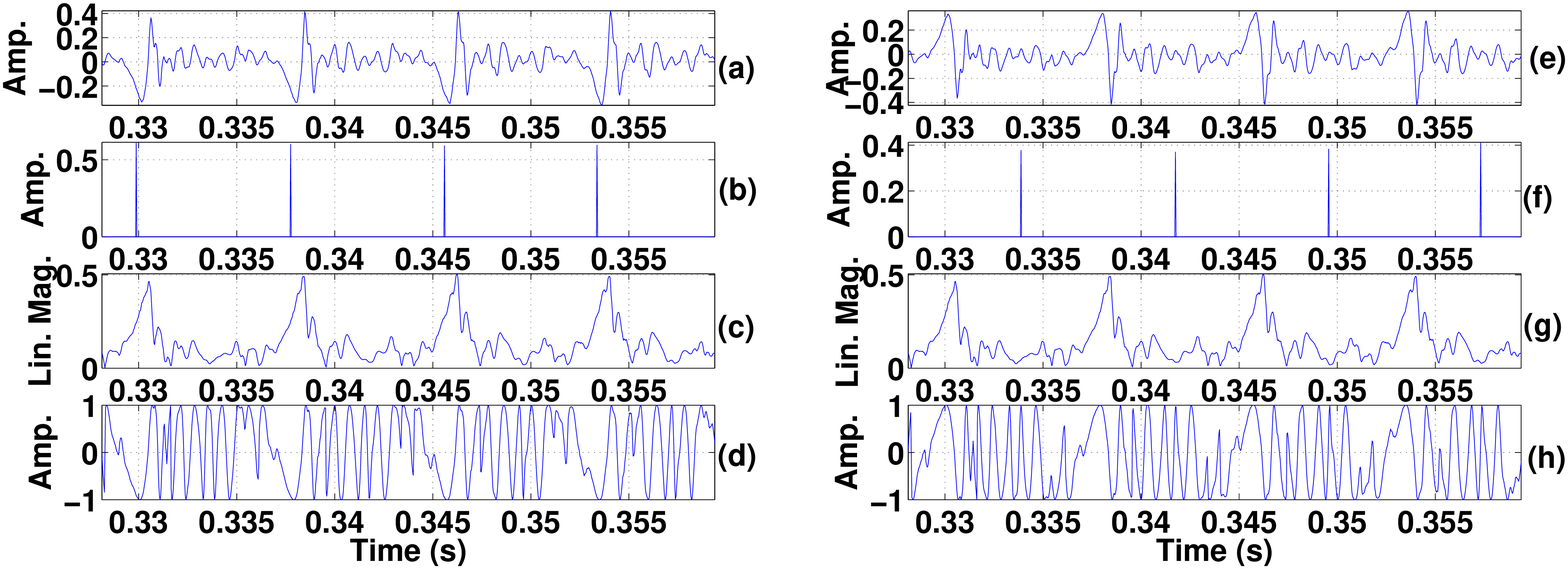,width=180mm,height=90mm}
\caption{ \label{speech_evidence}Variations in the cosine phase of analytic signal according to polarity changes in speech. A voiced segment of (a) speech , (b) instants of significant excitation (c) its HE and (d) cosine phase segment.
((e)-(h)) show the corresponding plots obtained when polarity of speech wave in (a) is reversed by multiplying  all the speech samples with -1. }
\end{figure*}

In the present work, the polarity is determined from the cosine phase of speech signal around the peaks in the HE of the speech signal. The HE peaks, nearest to the instants of significant excitation are selected for cosine phase analysis. The locations correspond to instants of significant excitation are estimated by the zero frequency filtering of speech signals as proposed by Murty et al.~\cite{Murty2008}. The slopes of the zero crossings in the cosine phase correspond to HE peak locations nearest to each instant location are then computed. Slopes of these zero crossings are found to be varying in accordance with the polarity of the signal.
\begin{figure*}
\epsfig{figure=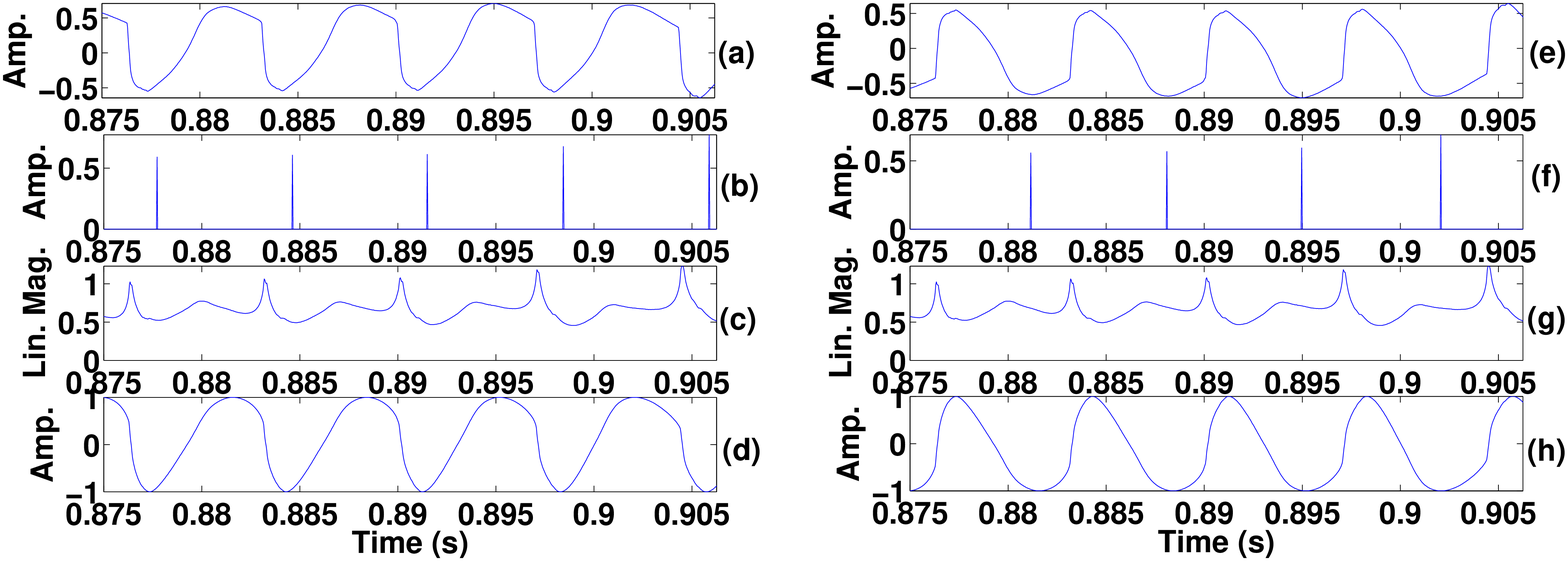,width=180mm,height=90mm}
\caption{ \label{egg_evidence}
Variations in the cosine phase of analytic signal according to polarity changes in EGG. A voiced segment of (a) glottal waveform , (b) corresponding estimated instants of significant excitation (c) its HE and (d) cosine phase segment. ((e)-(h)) show the corresponding plots obtained when polarity of glottal wave in (a) is reversed after multiplying all the EGG samples with -1. }
\end{figure*}
Left panel of the Figure~\ref{egg_evidence}((a)-(d))plot the EGG segment, estimated instants, HE of EGG segment and the corresponding cosine phase having positive polarity. The subplots in the right panel of the Figure~\ref{egg_evidence}((e)-(h)) are obtained when the polarity of waveform shown in Figure~\ref{egg_evidence}((a)) is inverted. Consistent phase patterns are observed in the regions of the cosine phase corresponds to HE peaks ((c)-(d)). Also, the characteristics of the phase patterns are reversed when the polarity of the given speech signal is inverted (compare the subplots (d) \& (h)). Hence the slopes of the cosine phase variation at the zero crossing correspond to the HE peaks are considered as the evidences of polarity of the signal. Polarity of the signal is finally decided based on the total number of positive or negative zero crossings of cosine phase at the HE peaks for the whole signal. In order to avoid the selection of spurious HE peaks, the peaks nearest to the estimated instants are considered. Also, another observation is that the HE of the signal remains unchanged irrespective of the polarity (subplots (c)\&(g)). Time shifts in the estimated instants can also be analyzed due to change in polarity of the signal (by comparing subplots (b)\&(f)). However, as long as HE peaks are invariant against polarity changes, there are no ambiguities in estimating the the nature of zero crossings in the cosine phase at the locations of the HE peaks. Figure~\ref{speech_evidence} demonstrates the the effect of polarity in estimated instants of significant locations, HE and cosine phase of a voiced segment of speech. Figure~\ref{speech_evidence}((d)\&(h)) show consistent phase variations in the cosine phase around the HE peak of the speech signal. The instants locations are also observed to be different for speech signals with opposite polarities.

%
%
\begin{table*}\center
\small
\caption{Performance evaluation of proposed polarity detection algorithm on CMU-Arctic database
and Emo-DB databases for speech.}
\label{Performance_Speech}
\scalebox{0.9}{
\begin{tabular}[b]{|c| c| c| c|c|c|c|}\hline
  & \multicolumn{3}{c|}{Proposed Method} & \multicolumn{3}{c|}{RESKEW Method}\\\cline{2-7}
 CMU Arctic Voice &\multicolumn{1}{c|}{No.of Corr. Polarity}&\multicolumn{1}{c|}{No. of False Polarity} & \multicolumn{1}{c|}{\% Correct}&\multicolumn{1}{c|}{No.of Corr. Polarity}&\multicolumn{1}{c|}{No. of False Polarity} & \multicolumn{1}{c|}{\% Correct}\\
 \hline
 $SLT$   &1130  & 0& \textbf{100} &1130  & 0& 100 \\\hline
 $BDL$ &1130  & 0&   \textbf{100}&1130& 0& 100\\\hline
 $JMK$   &1112  & 0&  \textbf{100} &1107 & 5& 99.55 \\\hline
 $CLB$ &1131  & 0&   \textbf{100}&1131 & 0& 100\\\hline
 $RSM$   &1131  & 0&  \textbf{100} &1131  & 0& 100 \\\hline
 $KSP$ &1131  & 0&   \textbf{100}&1131 & 0& 100\\\hline
 $AWB$   &1131  & 0&  \textbf{100} &1131 & 0& 100 \\\hline
 $EmoDb$&798&18&97.79&806&10&98.77\\ \hline
 $Total$&8695&18&99.79&8697&15&99.8 \\ \hline

\end{tabular}
}
\end{table*}
\section{Experimental Results}
\label{exp_results}
\subsection{Performance in clean conditions}

\begin{table*}\center
\small
\caption{Performance evaluation of proposed polarity detection algorithm on CMU-Arctic database
and Emo-DB databases for EGG.}
\label{Performance_egg}
\scalebox{0.9}{
\begin{tabular}[b]{|c| c| c| c|c|c|c|}\hline
  & \multicolumn{3}{c|}{Proposed Method} & \multicolumn{3}{c|}{RESKEW Method}\\\cline{2-7}
 CMU Arctic Voice &\multicolumn{1}{c|}{No.of Corr. Polarity}&\multicolumn{1}{c|}{No. of False Polarity} & \multicolumn{1}{c|}{\% Correct}&\multicolumn{1}{c|}{No.of Corr. Polarity}&\multicolumn{1}{c|}{No. of False Polarity} & \multicolumn{1}{c|}{\% Correct}\\
 \hline
 $SLT$   &1130  & 0& 100  &1130  & 0& 100 \\\hline
 $BDL$ &1130  & 0&   100&1130& 0& 100\\\hline
 $JMK$   &1113  & 0&  100 &1113 & 0& 100 \\\hline
 $EmoDb$ &789&27&96.69&796&20&97.55\\ \hline
 $Total$&4162&27&99.35&4169&20&99.52 \\ \hline

\end{tabular}
}
\end{table*}
The performance of the proposed polarity detection method is evaluated on two databases, namely, CMU-Arctic database and German emotional speech data base (EmoDb)~\cite{Kominek2004}~\cite{Burkhardt2005}. Both the databases are having simultaneous speech and EGG recordings. Seven synthesis specific speech corpora, each having 1131 phonetically balanced utterances and recorded at 32 kHz sampling rate, are used for the performance evaluation. Among the seven voice databases in CMU-Arctic used for the performance analysis, three voices (JMK,BDL \& SLT) are having simultaneous speech and EGG recordings. Hence, these three voices are used for performance evaluation of EGG data in CMU-Arctic database. The EGG data present in EmoDb is recorded with negative polarity as compared to the EGG data available in CMU-Arctic database. The EmoDb consists of seven emotions and with 10 texts spoken by 10 professional speakers per emotion. Each utterance in EmoDb is recorded at a sampling frequency of 16 kHz. The performance of the proposed polarity detection algorithm is measured based on the number of correctly and incorrectly classified instants for both speech and EGG. Table~\ref{Performance_Speech} and Table~\ref{Performance_egg} provide the comparative performance analysis of the proposed and RESKEW based polarity detection algorithms.

\begin{table*}\center
\small
\caption{Detection error rates of speech polarity detection under noise addition with babble noise at various SNR levels.}
\label{Noise_Performance}
\scalebox{0.9
}{
\begin{tabular}[b]{|c| c| c| c|c|c|c|}\hline
 SNR &0 (dB) & 5 (dB) &10 (dB) & 15 (dB) &20 (dB) & 30 (dB)\\
 \hline

 $Proposed$ & 0.012\%   &0.023\%  & 0.057 \%&  0.091\% &0.141\% & 0.232 \%\\\hline
 $RESKEW$ & 0.087\%   &0.121\%  & 0.114 \%&  0.114\% &0.137\% & 0.137 \%\\\hline
\end{tabular}
}
\end{table*}

Table~\ref{Performance_Speech} and ~\ref{Performance_egg} display the total detection error rates, 0.206\% and 0.644\% obtained for the proposed method from speech and EGG data respectively of the two databases. These results are comparable with the detection error rates of 0.172\% and 0.477\% obtained for RESKEW based polarity detection. Even though the performance of the proposed method is slightly less than that of the RESKEW method, the method strongly reinforces the significance of signal polarity information present in the cosine phase phase of the speech signals. For the reproducibility of the results, the codes that implements the proposed polarity detection algorithm is available for download at the following link: $http://nlp.amrita.edu:8080/TTS/polarityprograms.zip$.
\subsection{Performance in Noisy Conditions}
Table~\ref{Noise_Performance} shows the polarity detection error rates obtained for babble noise addition at the various SNR levels starting from 0 dB. As the discontinuities at the instants of significant excitation are well preserved for different levels of noise addition in the speech, the polarity detection error rates are least affected by the noise. Also, the noise robustness of proposed method is almost similar to that obtained for the RESKEW method.

\section{Summary \& Conclusion}
\label{summary}
In this paper, we propose a method to detect the polarity using cosine phase of the signals. The proposed method relies on the estimation of instants of significant excitation for selecting the relevant peaks in the HE  for detecting the polarity of the speech or EGG. The signal polarity is  then determined by analyzing the  nature of the zero crossings of cosine phase around the selected HE peaks. The effectiveness of the proposed method is demonstrated in terms of reduced detection error rates. Also, the performance of the proposed polarity detection is almost equal or slightly less than that of the existing RESKEW method for polarity detection. Since the cosine phase analysis is anchored around instants of significant excitation, the noise performance of the proposed method is also found to be robust.

\section{Acknowledgements}
The present work is a part of ongoing DST-Fastrack project titled "Analysis, processing and synthesis of emotions in speech" carried out at Amrita Vishwa Vidyapeetham, Coimbatore, Tamilnadu. The funding agency of the project is SERC, New Delhi.

\bibliographystyle{IEEEtran}
\bibliography{references}

\begin{thebibliography}{1}
\providecommand{\url}[1]{#1}
\csname url@samestyle\endcsname
\providecommand{\newblock}{\relax}
\providecommand{\bibinfo}[2]{#2}
\providecommand{\BIBentrySTDinterwordspacing}{\spaceskip=0pt\relax}
\providecommand{\BIBentryALTinterwordstretchfactor}{4}
\providecommand{\BIBentryALTinterwordspacing}{\spaceskip=\fontdimen2\font plus
\BIBentryALTinterwordstretchfactor\fontdimen3\font minus
  \fontdimen4\font\relax}
\providecommand{\BIBforeignlanguage}[2]{{%
\expandafter\ifx\csname l@#1\endcsname\relax
\typeout{** WARNING: IEEEtran.bst: No hyphenation pattern has been}%
\typeout{** loaded for the language `#1'. Using the pattern for}%
\typeout{** the default language instead.}%
\else
\language=\csname l@#1\endcsname
\fi
#2}}
\providecommand{\BIBdecl}{\relax}
\BIBdecl

\bibitem{Dugman2011}
T.~Dugman and T.~Dutoit, ``Oscillating statistical moments for speech polarity
  detection,'' in \emph{Non-{L}inear {S}peech {P}rocessing {W}orkshop
  (NOLISP11)}, 2011, pp. 48--54.

\bibitem{Ding1998}
W.~Ding and N.~Campbell, ``Determining polarity of speech signals based on
  gradient of spurious glottal waveform,'' in \emph{{ICASSP 98}}, 1998, pp.
  857--859.

\bibitem{Drugman2013}
T.~Drugman, ``Residual excitation skewness for automatic speech polarity
  detection,'' \emph{IEEE Signal Process. Letters}, vol.~20, no.~4, pp.
  387--390, April 2013.

\bibitem{Saratxaga2009}
I.~Saratxaga, D.~Erro, I.~Hernáez, I.~Sainz, and E.~Navas, ``Use of harmonic
  phase information for polarity detection in speech signals,'' in
  \emph{{INTERSPEECH 2009}}, 2009, pp. 1075--1078.

\bibitem{Murty2008}
K.~S.~R. Murty and B.~Yegnanarayana, ``Epoch extraction from speech signals,''
  \emph{IEEE Trans. Audio, Speech and Language Process.}, vol.~16, no.~8, pp.
  1602--1614, Nov. 2008.

\bibitem{Cohen1995}
L.~Cohen, \emph{Time-Frequency Analysis: Theory and Applications}, S.~P.
  Series, Ed.\hskip 1em plus 0.5em minus 0.4em\relax ser. Signal Processing
  Series. Englewood Cliffs: Prentice-Hall, 1995.

\bibitem{Kominek2004}
J.~Kominek and A.~Black, ``{CMU}-{A}rctic speech databases,'' in \emph{in 5th
  ISCA Speech Synthesis Workshop}, Pittsburgh, PA, 2004, pp. 223--224.

\bibitem{Burkhardt2005}
F.~Burkhardt, A.~Paeschke, M.~Rolfes, W.~Sendlemeier, and B.~Weiss, ``A
  database of {G}erman emotional speech,'' in \emph{Proc. INTERSPEECH}, 2005,
  pp. 1517--1520.

\end{thebibliography}

\end{document}